\documentclass[aip,jcp,reprint]{revtex4-1}

\usepackage[ngerman, english]{babel}
\usepackage{graphicx}
\usepackage{dcolumn}%
\usepackage{bm}%
\usepackage{geometry}
\usepackage{array}
\usepackage{amsthm}
\usepackage{amstext}
\usepackage{amsmath,amssymb}
\usepackage{nicefrac,subfigure}
\usepackage{amsopn}

\def\({\left(} \def\){\right)}

   \def\lw{\left\langle} \def\rw{\right\rangle}

     \let \e=\varepsilon
  \let\o=\omega  \let\s=\sigma

\newcommand{\tr}{^{\text{\text{T}}}}

\def\Id{{\rm Id}}
\def\op{{\rm op}}

\def\tr{{\rm tr}}

\def\C{\mathbb C}
\def\N{\mathbb N}

\def\R{\mathbb R}

\begin{document}

\title[Quasi-classical molecular dynamics by Egorov's theorem]{Quasi-classical description of molecular dynamics based on Egorov's theorem}
\author{Johannes Keller}
\email[]{keller@ma.tum.de}
\affiliation{Zentrum Mathematik, Technische Universit\"at M\"unchen, 80290
M\"unchen, Germany}
\author{Caroline Lasser}
\email[]{classer@ma.tum.de}
\affiliation{Zentrum Mathematik, Technische Universit\"at M\"unchen, 80290
M\"unchen, Germany}

\date{\today}

\begin{abstract}
Egorov's theorem on the classical propagation of quantum observables is related to prominent quasi-classical descriptions of quantum molecuar dynamics
as the linearized semiclassical initial value representation (LSC-IVR), the Wigner phase space method or the statistical quasiclassical method. 
The error estimates show that different accuracies are achievable for the computation of expectation values and position densities. 
Numerical experiments for a Morse model of diatomic iodine and confined Henon--Heiles systems in various dimensions illustrate the theoretical results.  
\end{abstract}


\maketitle


\section{Introduction}

The numerical simulation of quantum molecular dynamics is a notoriously difficult problem, since the key equation, the vibrational time-dependent Schr\"odinger equation, is a partial differential equation on a high dimensional configuration space with solutions, that oscillate in time and space. 

Over decades this challenge has been tackled by methods that directly compute quantities of physical interest without solving the Schr\"odinger equation 
or fully discretizing its unitary propagator. The linearized semiclassical initial value representation (LSC-IVR)\cite{M74,WSM98,TW04}, for example, approximates time-dependent correlation functions and expectation values by initial phase space sampling and classical trajectory calculations.  The Wigner phase space method\cite{He76,BH81,DH84} and the statistical quasiclassical method \cite{LS80} similarly approximate time-dependent transition probabilities. 

A unifying property of these quasi-classical approaches is  the following three-step procedure:
(i) Sampling of an initial phase space density
(ii) Classical propagation of the sampling points
(iii) Weighted summation over the time-evolved phase space points. 
Notably the second and third algorithmic step are numerically more favorable than solving the time-dependent Schr\"odinger equation in higher dimensions. Often the computational times are in the range of seconds.

Quasi-classical methods are well-established in the literature and have been thoroughly discussed also with respect to deficiencies for quantum coherence on longer time scales \cite{SWM98,TW04} or zero point energy leakage\cite{HM09}. They have been derived from the asymptotic expansion of the Wigner transformed Schr\"odinger equation\cite{He76}, semiclassical initial value representations\cite{SWM98} and the path integral formulation\cite{SG03,PNR03} of the unitary propagator. 

Our aim here is to add a complementary derivation by relating quasi-classical algorithms to Egorov's theorem\cite{E69,BR02} on the classical propagation of quantum observables. Moreover, Egorov's theorem also implies error estimates for the computation of time-evolved expectation values and position densities.
In all cases, the error crucially depends on the time evolution of derivatives of the classical trajectories with respect to their initial data. But more can be inferred: One assumes that the vibrational Schr\"odinger operator can be written as
\begin{equation}\label{eq:hamiltonian}
\widehat H = -\frac{\e^2}{2}\Delta + V,
\end{equation}
where $\e$ is a small positive parameter and $V$ a potential energy surface (PES). Then, time-dependent expectation values are approximated with an error of the order $\e^2$ for all initial states with $\langle\psi|\psi\rangle=1$. The approximation of position densities and transition probabilites, however, requires localization assumptions on the initial state, and in typical vibrational situations one can only expect an error of the order $\sqrt\e$. 

We proceed as follows: In \S\ref{sec:vqd} we present Egorov's theorem together with estimates for the time-dependance of the error. In \S\ref{sec:cm} we relate Egorov's theorem to the linearized semiclassical initial value representation (LSC-IVR) and the Wigner phase space method. \S\ref{sec:three-prop} discusses the computational tasks of quasi-classical algorithms. In \S\ref{sec:ne} we present numerical experiments for a Morse-model of diatomic Iodine and confined Henon--Heiles systems ranging from dimension $2$ to $32$. \S\ref{sec:con} summarizes our results, while the Appendices collect elements of our theoretical error analysis.

\section{Vibrational Quantum Dynamics}\label{sec:vqd}

\subsection{Unitary propagator}

Within the framework of the Born--Oppenheimer approximation, the Schr\"odinger operator for effective 
nuclear dynamics related to a single electronic state writes in atomic units as 
$$
\widehat H = -\sum_{j=1}^d\frac{1}{2m_j}\Delta_{q_j} + V,
$$
where $m_j$ is the mass for the $j$th component of the nuclear coordinate vector. The real-valued function $V:\R^d\to\R$ is a potential energy surface (PES) of the molecular system. 

Setting 
$$
\e = 1/\sqrt{\max(m_1,\ldots,m_d)}
$$
and scaling the coordinates according to $q_j\mapsto q_j/(\e\sqrt{m_j})$, we write the Schr\"odinger operator in the semiclassical form (\ref{eq:hamiltonian}) and study the vibrational dynamics on the long time scale $t/\e$, that is, we use the time-scaled unitary propagator
$$
U_t = e^{-i\widehat Ht/\e}.
$$
Depending on the nuclear masses, the scale parameter $\e$ ranges between $10^{-3}$ and $10^{-2}$. For example, the diatomic iodine molecule considered later on has $\e=0.0122$, and accordingly one unit of the long time scale corresponds  to $1.98$ femtoseonds.

\subsection{Observables}

The observables result from the Weyl quantization of functions $A:\R^d\times\R^d\to\C$ according to
\begin{eqnarray*}
\lefteqn{(\widehat A\psi)(q) =}\\
&& (2\pi\e)^{-d} \int A(\tfrac12(q+y),p) e^{i(q-y)\cdot p/\e} \psi(y) dp dy,
\end{eqnarray*}
where $\psi:\R^d\to\C$ is a square-integrable function. The $\e$-scaling of the Fourier term allows to view the Schr\"odinger operator 
$\widehat H$ as the quantization of the $\e$-independent energy function
\begin{equation}\label{eq:ham}
H(q,p) = \tfrac12 |p|^2 + V(q).
\end{equation}
Also the position and momentum operators $\psi\mapsto q_j\psi$ and $\psi\mapsto -i\e\partial_j\psi$ for $j=1,\ldots,d$ originate from the $\e$-independent phase space functions
$(q,p)\mapsto q_j$ and $(q,p)\mapsto p_j$, respectively. 

For the trace of two Weyl quantized observables one has the beautiful integral formula 
$$
\tr(\widehat A \widehat B) = (2\pi\e)^{-d} \int A(z) B(z) dz.
$$

\subsection{Wigner functions}
Expectation values for Weyl quantized observables can be expressed in terms of the Wigner function $W_\psi:\R^d\times\R^d\to\R$, 
\begin{eqnarray*}
\lefteqn{W_\psi(q,p) =}\\
 &&(2\pi\e)^{-d} \int \psi(q-\tfrac12y)\psi^*(q+\tfrac12y) e^{i p\cdot y/\e} dy,
\end{eqnarray*}
via
$$
\langle \psi \mid \widehat A\mid \psi \rangle = \int A(z) W_\psi(z) dz.
$$
Moreover, the Weyl quantization of the Wigner function $W_\psi$ is the projector for $\psi$,
$$
(2\pi\e)^d\; \widehat{W_\psi} = |\psi\rangle \langle \psi|.
$$

A typical initial state for vibrational quantum dynamics is the ground state of an harmonic oscillator 
$\widehat H = -\tfrac{\e^2}{2}\Delta + \tfrac12 |q|^2$, 
or slightly more general, a localized Gaussian wavepacket with phase space center $z_0=(q_0,p_0)$,
\begin{align}
\psi &\nonumber (q) = (\det(\Sigma)/\pi\e)^{d/4} \exp\!\( \tfrac i\e p_0\cdot(q-q_0)\)\\
&\label{eq:coherent_state}\times \exp\!\( - \tfrac1{2\e}(q-q_0) \cdot\Sigma (q-q_0) \),
\end{align}
where $\Sigma\in\R^{d\times d}$ is a positive definite diagonal matrix
with entries $\s_1,\ldots,\s_d$. Its Wigner function is given by
\begin{align}
\lefteqn{W_\psi(z) =}\nonumber\\
&&\label{eq:wigner_gauss} (\pi\e)^{-d}\exp\!\(-\tfrac{1}{\e} (z-z_0)\cdot \Sigma_2 (z-z_0)\),
\end{align}
where $\Sigma_2\in\R^{2d\times 2d}$ is the diagonal matrix with diagonal entries $\sigma_1,\ldots,\sigma_d,1/\sigma_1,\ldots,1/\sigma_d$.  

In contrast to the Gaussian wavepacket~(\ref{eq:coherent_state}), most Wigner functions attain negative values. The Wigner functions of the Hagedorn wavepackets or the generalized squeezed states, for example, can be expressed as the product of a Gaussian and a Laguerre polynomial\cite{LT14}. 
In general, however, analytical formulas are not available, and Wigner functions have to be computed numerically, which poses a very difficult problem of high-dimensional oscillatory numerical integration.

\subsection{Egorov's theorem}\label{sec:repres}

Quasi-classical approximations rely on the flow 
$\Phi_t:\R^{2d}\to\R^{2d}$ of the classical Hamiltonian function~(\ref{eq:ham}). The flow relates  initial phase space points $(q_0,p_0)$ with their location at time~$t$. One has $\Phi_t(q_0,p_0)=(q_t,p_t)$ with
\begin{align}\label{eq:hamilton_equation}
\dot{q_t} = p_t,\qquad \dot{p_t} = -\nabla V(q_t).
\end{align}
Egorov's theorem\cite{E69,BR02} proves for the propagation of Weyl quantized observables that
\begin{equation}\label{egorov}
U_{-t} \,\widehat B\, U_t = \widehat{B\circ\Phi_t} + \e^2 E(V, B,\Phi_t)
\end{equation}
holds, where the error term $E(V, B, \Phi_t)$ depends on the following:
\begin{enumerate}
\item[(i)] potential derivatives $\partial^\alpha V$ with $|\alpha|\ge 3$, 
\item[(ii)] observable-flow derivatives $\partial^\alpha (B\circ\Phi_t)$ with $|\alpha|\ge 1$,
\end{enumerate}
see Appendix~\ref{proof}. If the potential $V$ is a polynomial of degree less or equal than two, then $E(V,B,\Phi_t)=0$, and the classical propagation 
of observables exactly describes the quantum evolution. Moreover, if $B=H$, then $E(V,H,\Phi_t)=0$ as well.


\subsection{Ehrenfest time}\label{sec:wigner-egorov}

For the analysis of Egorov's theorem (cf. Ref.~\cite{BR02} and Appendix~\ref{proof}), the derivatives of the classical Hamiltonian flow are crucial.  The worst case estimate gives for any multi-index $\alpha\in\N^{2d}$ a constant $C_\alpha>0$ such that for all $t\in\R$ and $z\in\R^{2d}$
\begin{equation}\label{exponential}
\mid \partial^\alpha \Phi_t(z)\mid \le C_\alpha e^{\Gamma |\alpha| \cdot |t|},
\end{equation}
where the flows's stability indicator $\Gamma>0$ is related to the eigenvalues of the Hessian matrix~$D^2V(q)$ of the potential. 

The worst case exponential growth of the flow derivatives (\ref{exponential}) implies exponential growth of the error in Egorov's theorem, a phenomenon, which is well-established for nonsymmetric double well potentials\cite{BR02}. Hence, in the worst case, one has to expect that the $\e^2$ factor in (\ref{egorov}) is consumed after times $t$ of the order $\log(1/\e)$, the so-called Ehrenfest time scale.

For integrable systems or flows with closed orbits, the exponential estimate (\ref{exponential}) can be relaxed to
$$
\mid \partial^\alpha \Phi_t(z)\mid \le C_\alpha (1+|t|)^\alpha,
$$
and Egorov's approximation is meaningful until times of the order $1/\sqrt\e$, see Ref.\cite{BR02}. Our numerical experiments for a model of diatomic iodine and a modified Henon--Heiles system even show persistence on longer time scales.

\section{Computational methods}\label{sec:cm}

Over decades, quasiclassical approximations in the spirit Egorov's theorem have been used as the backbone for numerical methods in molecular quantum dynamics. We exemplarily summarize two of them.

\subsection{LSC-IVR}\label{sec:LSC}

The linearized semiclassical initial value representation (LSC-IVR)\cite{M74,WSM98,TW04} approximates time-dependent correlation functions by
$$
\tr\!\left(\widehat A\, U_{-t}\widehat B U_t\right) \approx (2\pi\e)^{-d} \int A(z) B(\Phi_t(z)) \,dz,
$$
and in particular
\begin{eqnarray*}
\tr\!\left( |\psi\rangle\langle\psi| \,U_{-t}\widehat B U_t\right) 
&=& \langle U_t\psi\mid \widehat B \mid U_t\psi\rangle\\
&\approx &\int W_\psi(z) B(\Phi_t(z)) \,dz.
\end{eqnarray*}
In the literature, LSC-IVR is derived from semiclassical initial value representations of the unitary propagator $U_t$. However, Egorov's theorem offers a simpler proof.

According to the Egorov estimate~(\ref{egorov}), the LSC-IVR approximates quantum correlation functions and expectation values with an error of the order~$\e^2$, if the initial state is normalized to $\langle\psi|\psi\rangle=1$ , and if the time-evolved observable originates from Weyl quantizing an $\e$-independent phase space function $B$ with bounded derivatives\cite{LR10}. 

For the approximation of time-evolved position densities, one writes\cite{SM99}
\begin{align}
\lefteqn{| (U_t\psi)(r)|^2 \;=\; \langle U_t\psi\mid \delta_r \mid U_t\psi\rangle} \nonumber\\
&= (2\pi)^{-d} \int \langle U_t \psi \mid e^{-i\eta\cdot(q -r)} \mid U_t\psi\rangle d\eta \nonumber\\
&\approx\label{eq:pos_dens_app}  \int W_\psi(z) B_\eta(\Phi_t(z)) dz d\eta 
\end{align}
with $B_\eta(q,p)=(2\pi)^{-d}e^{-i\eta\cdot(q-r)}$. 

For this approximation of $|(U_t\psi)(r)|^2$ , the accuracy crucially depends on the initial state~$\psi$. For a vibrational Gaussian wavepacket (\ref{eq:coherent_state}), for example, the approximation error is of the order~$\sqrt\e$, see Appendices~\ref{pd} and \ref{app:heur}.

\subsection{Wigner phase space method}

The Wigner phase space method\cite{He76,BH81,DH84} and the statistical quasiclassical method \cite{LS80} approximate time-dependent transition probabilities as
\begin{eqnarray*}
\mid \langle \phi\mid U_t\psi\rangle\mid^2 &=& \langle U_t\psi\mid \phi\rangle \langle\phi\mid U_t\psi\rangle\\
&\approx & (2\pi\e)^d \int W_\phi(\Phi_t(z)) W_\psi(z) dz.
\end{eqnarray*}
Here, Egorov's theorem is used in the form
$$
U_{-t}\;|\phi\rangle\langle\phi| \;U_t \;\approx\; \widehat{B\circ\Phi_t}
$$
with $B = (2\pi\e)^d \,W_\phi$, that is, $\widehat B = |\phi\rangle\langle\phi|$. Hence, the Wigner phase space method is a special case of LSC-IVR, though typically derived from asymptotic expansions of the Wigner function.

The accuracy of this method depends on the states $\phi$ and $\psi$. 
If they are vibrational states, as for example localized Gaussian wavepackets as defined in Eq.~(\ref{eq:coherent_state}), then the third derivatives of the Wigner function contribute terms of the order $\e^{-3/2}$, such that the overall approximation error is of the order $\e^{2-3/2} = \e^{1/2}$, see Appendix~\ref{app:heur}.

\section{Computational tasks}\label{sec:three-prop}


For the quasi-classical approximation of expectation values
$$
\langle U_t\psi\mid \widehat B\mid U_t\psi\rangle \approx \int W_\psi(z) B(\Phi_t(z)) dz,
$$
the following three computational steps have to be carried out:

(i) Sampling of the initial condition: We choose phase space points $(q_1,p_1),\ldots,(q_{N},p_{N})$ such that
\begin{equation}\label{eq:quad_initial}
\langle \psi\mid \widehat B \mid\psi\rangle \approx \frac{1}{N} \sum_{j=1}^{N} B(q_j,p_j) 
\end{equation}
for the observables $B$ of interest. This is achieved by Monte Carlo or Quasi-Monte Carlo sampling of the initial Wigner function $W_{\psi}$. If the Wigner function $W_\psi$ attains negative values, one can apply stratified or importance sampling\cite{LR10}.
We note that an unrefined sampling of the initial Husimi function deteriorates the accuracy of the algorithm\cite{KLW09,KL13}.

(ii) Classical trajectory calculations: The chosen phase space points are evolved along the trajectories of the corresponding classical Hamiltonian 
system
$$
\dot q_t = p_t,\qquad \dot p_t = -\nabla V(q_t).
$$
Since the observables of interest are computed by phase space averaging, these classical equations of motion should be discretized symplectically as e.g. by the St\"ormer--Verlet method or by higher order symplectic 
Runge--Kutta schemes, see \ref{sec:symplectic}.

(iii) Evaluation of the observables: At some time~$t$, the algorithm has resulted in phase space points $(q_1(t),p_1(t)),\ldots,(q_{N}(t),p_{N}(t))$. Then, the expectation values of interest are approximated according to
\begin{equation}\label{eq:quadrature_expectation}
\langle U_t \psi\mid \widehat B \mid U_t\psi\rangle \approx \frac1N\sum_{j=1}^{N} B(q_j(t),p_j(t)). 
\end{equation}

\subsection{Phase space sampling and quadrature}\label{sec:sampling}

We discuss the initial sampling step for Gaussian wave packets of the form~(\ref{eq:coherent_state}). 
Monte Carlo samplings of the corresponding phase space Gaussian~(\ref{eq:wigner_gauss}) can easily be generated by a suitably 
rescaled and shifted sampling of a standard $2d$-dimensional Gaussian distribution.
The convergence rate of the Monte-Carlo quadrature rule~(\ref{eq:quad_initial}) is 
proportional to $1/\sqrt{N}$, where $N$ is the number of sampling points. 
Quasi-Monte Carlo sequences, such as Sobol or Halton sequences,  approximate 
the uniform distribution on the unit cube. To obtain a Gaussian distribution with diagonal covariance matrix, one transforms the
uniformly distributed sequences by the cumulative distribution functions of $2d$ univariate Gaussians. 
The rate of convergence for Quasi-Monte Carlo quadratures is approximately 
given by\cite{LR10}  $\log(N)^{2d}/N$, and hence detoriates slightly with increasing dimension.

\begin{figure}[ht!]
\includegraphics[width=7.5cm]{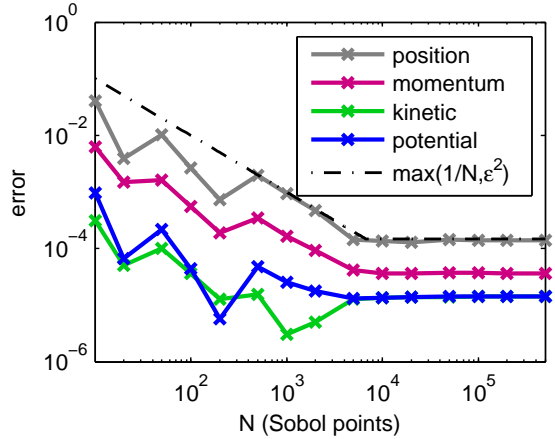}
\caption{Average errors on the time interval $[0,166{\rm fs}]$ for the one-dimensional Morse
system from \S\ref{sec:numerics_iodine} and different numbers $N$ of Sobol points.}\label{fig:points_convergence}
\end{figure}
Figure~\ref{fig:points_convergence} illustrates  the numerical convergence of the Sobol quadrature 
rule~(\ref{eq:quadrature_expectation}) when applied to the one-dimensional Morse system from~\S\ref{sec:numerics_iodine}. 
The errors are averaged over the time interval $[0,166{\rm fs}]$, and we used the St\"ormer-Verlet scheme with
stepsize $\tau=10^{-3}$ for the dynamics. One observes that the quadrature error is bounded by the maximum of $1/N$ and $\e^2$, 
the $1/N$ error originating from the Quasi-Monte Carlo quadrature, the $\e^2$ error originating from the asymptotic approximation of 
Egorov's theorem.

\subsection{Propagation with symplectic integrators}\label{sec:symplectic}

\begin{figure}[ht!]
\includegraphics[width=7.5cm]{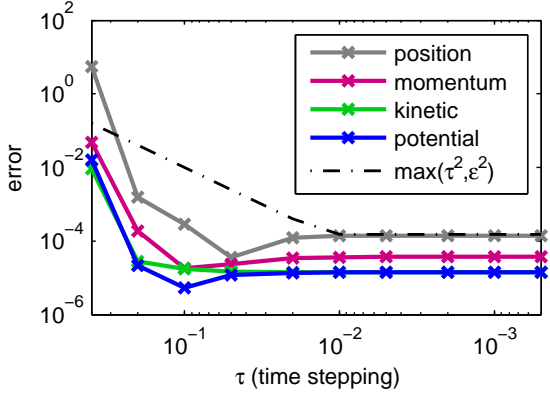}
\caption{Average errors on the time interval $[0,{\rm 166fs}]$ for the one-dimensional Morse
system from \S\ref{sec:numerics_iodine} with different time steppings $\tau$ for the St\"ormer-Verlet
 integrator.}\label{fig:steps_convergence}
\end{figure}
The most popular symplectic integrator is the St\"ormer-Verlet scheme which is a symmetric 
second order method. Its application to the Hamiltonian system~(\ref{eq:hamilton_equation}) with time stepping $\tau$
results in the  update formula $(q_n,p_n)\mapsto (q_{n+1},p_{n+1})$ with
\begin{align}
q_{n+1/2} &\nonumber= q_n + \tfrac \tau 2 p_n\\
p_{n+1} &\nonumber= p_n - \nabla V(q_{n+1/2})\\
q_{n+1} &\label{eq:stoermer-verlet} = q_{n+1/2} + \tfrac \tau 2 p_{n+1}.
\end{align}
Higher order symplectic integrators\cite{Y90,HLW10} can be constructed by a similar splitting procedure.
Figure~\ref{fig:steps_convergence} shows the second order accuracy of the St\"ormer-Verlet
scheme applied to the Morse oscillator from~\S\ref{sec:numerics_iodine}. We used $10^5$ Sobol points for the
Quasi-Monte Carlo quadrature.
 Already for moderately small time steppings $\tau$, the $\e^2$ error from the Egorov theorem is dominant.


\subsection{Evaluation of position densities}\label{sec:wigner_position_dens}

For the approximation of position densities according to
\begin{align}
|(U_t\psi)(r)|^2 &\approx \nonumber (2\pi)^{-d} \int W_\psi(z) e^{-i\eta\cdot(q_t-r)} dz d\eta\\
&\label{eq:egorov_pos_dens}=: P_t(r),
\end{align}
the previous algorithmic steps (i)--(iii), have to be augmented by an additional quadrature step. This step is, however, only feasible for low dimensional systems:

(iv) Evaluation of the position density: We choose quadrature nodes $\eta_1,\ldots,\eta_M$ and weights $w_1,\ldots,w_M$ such that
$$
|(U_t\psi)(r) |^2 \approx \frac{1}{(2\pi)^d N} \sum_{j=1}^N \sum_{k=1}^M e^{-i \eta_k \cdot (q_j(t)-r)} w_k
$$
This can be achieved by the Fast Fourier Transform (FFT), since the $\eta$-integral defining $P_t(r)$ is an inverse Fourier transform.


\section{Numerical Experiments}\label{sec:ne}

All the computations presented in this chapter have been performed 
with \textsc{Matlab} $8.3$ on a $3.33$GHz Intel Xeon X5680 processor. 
The algorithmic structure suggests parallel and GPU computing. 
Preliminary tests in this direction indicate considerable speed-ups. 

\subsection{Ground state dynamics for diatomic Iodine}\label{sec:numerics_iodine}

We first present simulations for the dynamics
of a diatomic iodine molecule on the lowest potential energy surface, that is, the electronic ground state of $I_2$.

\subsubsection{The model system}
The vibrational degree of freedom is the internuclear distance $r$, and the electronic ground
state energy is modelled by a Morse potential fitted to experimental 
data\cite{BY73},
\begin{equation}\label{eq:morse}
V_{I_2}(r) = D_e (1-e^{-\alpha (r - r_e)})^2
\end{equation}
with $D_e = 0.0572$ hartree, $\alpha = 0.983 a_0^{-1}$, and $r_e = 5.03855 a_0$, where the Bohr radius~$a_0$ is unity in atomic units.
The associated Schr\"odinger Hamiltonian
$$
\widehat H = -\frac{1}{2m}\partial_r^2 + V_{I_2}
$$
with reduced mass parameter $m=1.165\cdot 10^5$ a.u. has previously been used in the literature\cite{FM96,WTSGGM01,TW04}.

To identify the effective semiclassical scale of this model, we set the energy unit to $D_e$, which yields
the rescaled Hamiltonian
 $$
\widehat H = -\frac{ \e^2}{2}\partial_r^2 + (1-e^{-\alpha (r - r_e)})^2
$$
with $ \e = \sqrt{1/(m D_e)} = 0.0122$ and the corresponding Schr\"odinger equation
\begin{equation}\label{eq:morsehamil}
i\e \partial_t \psi(r,t) =\widehat H\psi(r,t).
\end{equation}

As the initial state we consider a one-dimensional Gaussian wave packet~(\ref{eq:coherent_state})
with width parameter $\Sigma=1.3836$
and phase space center $(q_0,p_0) = (4.53,0)$, 
which corresponds to the initial data 
previously used for the analysis of a forward-backward
IVR method\cite{WTSGGM01} with the same potential.


\subsubsection{The numerical setup}

The references solutions for the Schr\"odinger equation~(\ref{eq:morsehamil}) are obtained by a high resolution Fourier split-step method with computational parameters listed in Table~\ref{tab_ref1}. The final time $1668$fs corresponds to roughly $836$ time units with respect to the macroscopic time scale $t/\e$. 
\begin{table}[ht!]
\centering
\begin{tabular}{c|c|c|c}
$r$ interval& Fourier modes &  time  & timesteps\\ \hline
 $[3,11]$ & $2\cdot 10^4$ & $[0,1668{\rm fs}]$  &  $2\cdot 10^6$ \\
\end{tabular}
\caption{\label{tab_ref1} Data of the reference solution for the 
vibrational Schr\"odinger equation~(\ref{eq:morsehamil}) with $\e=0.0122$.}
\end{table}

For the quasiclassical computation of expectation values, we sample the initial Wigner function with $10^5$ Monte Carlo points, and
perform the propagation with a time stepping $\tau=4\cdot10^{-3}$ for the St\"ormer-Verlet integrator, see~\S\ref{sec:symplectic}. 
Then we take the mean over ten independent runs of this setup.

For the computation of position densities according to \S\ref{sec:wigner_position_dens}, we use $4\cdot10^5$ Monte Carlo points in ten independet runs,
a symplectic integrator\cite{Y90} of order eight with time stepping $\tau=10^{-2}$,
and $2^{12}$ Fourier modes for the inverse Fourier transform.

\begin{figure}[h!]
\includegraphics[width=7.5cm]{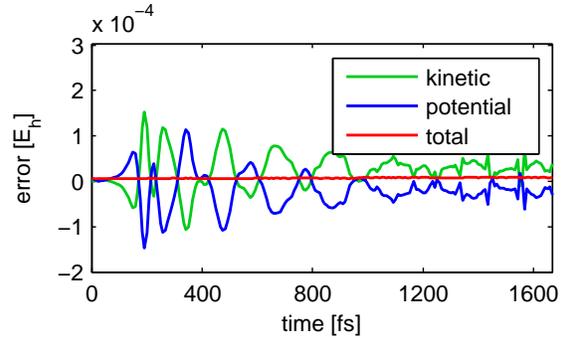}
\caption{Evolution of the differences between the expected energies computed by the quasi-classical algorithm
for the Iodine potential~(\ref{eq:morse}) and references obtained from highly accurate quantum mechanical calculations, 
see Table \ref{tab_ref1}.}\label{fig:morse_error}
\end{figure}

\subsubsection{Expectation values}
\begin{figure}[h!]
\includegraphics[width=7.5cm]{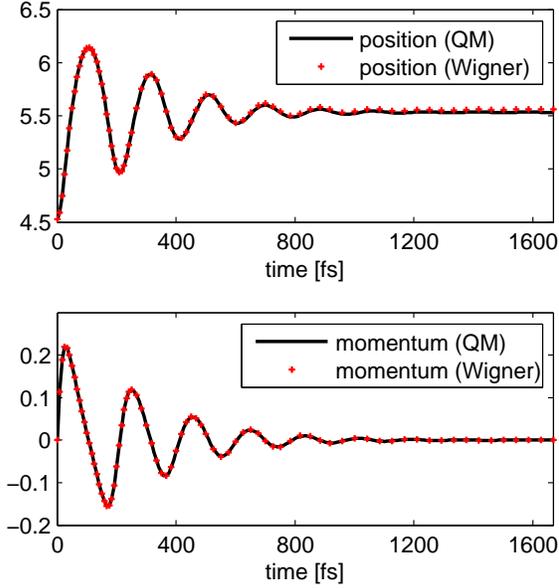}
\caption{Evolution of the expected position and momentum in the $I_2$ system 
for both, the quasi-classical algorithm and quantum mechanical calculations.}\label{fig:morse_posmom}
\end{figure}

The evolution of the kinetic, potential, and total energy errors from our numerical experiments is shown in Figure~\ref{fig:morse_error}. It illustrates
 total energy conservation of the quasi-classical algorithm and shows small kinetic and potential energy errors over long times. 
Also for the evolution of the position and momentum expectation, the results of the quasi-classical algorithm and the quantum mechanical references are very close, see Figure~\ref{fig:morse_posmom}.

\begin{figure}[h!]
\includegraphics[width=7.5cm]{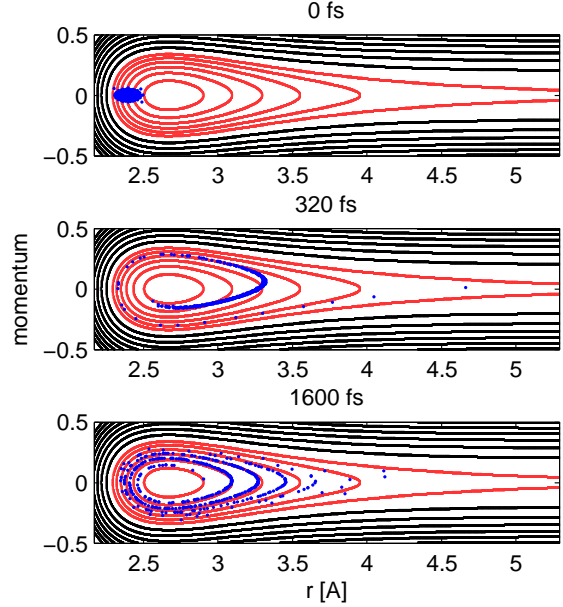}
\caption{The dynamics of 500 initial sampling points superimposed on selected contour lines of the classical total energy. The red and black contours correspond to the trapping respectively unbounded energy region.}\label{fig:morse_portrait}
\end{figure}

In our simulations almost all of the classical trajectories are trapped in the Morse well,  
since the initial state is localized in the potential well with small kinetic energy. $99.57\%$
of the Sobol points generated for the initial data lie within the trapping region, see the blue dots on top of the red contour lines in Figure~\ref{fig:morse_portrait}.
The stability and periodicity of the classical flow in this region imply that the error estimates of Egorov's theorem 
stay small up to times much longer than the uniform Ehrenfest timescale, see~\S\ref{sec:wigner-egorov}.

\subsubsection{Position densities}\label{sec:posdens}
\begin{figure*}
\includegraphics[width=15cm]{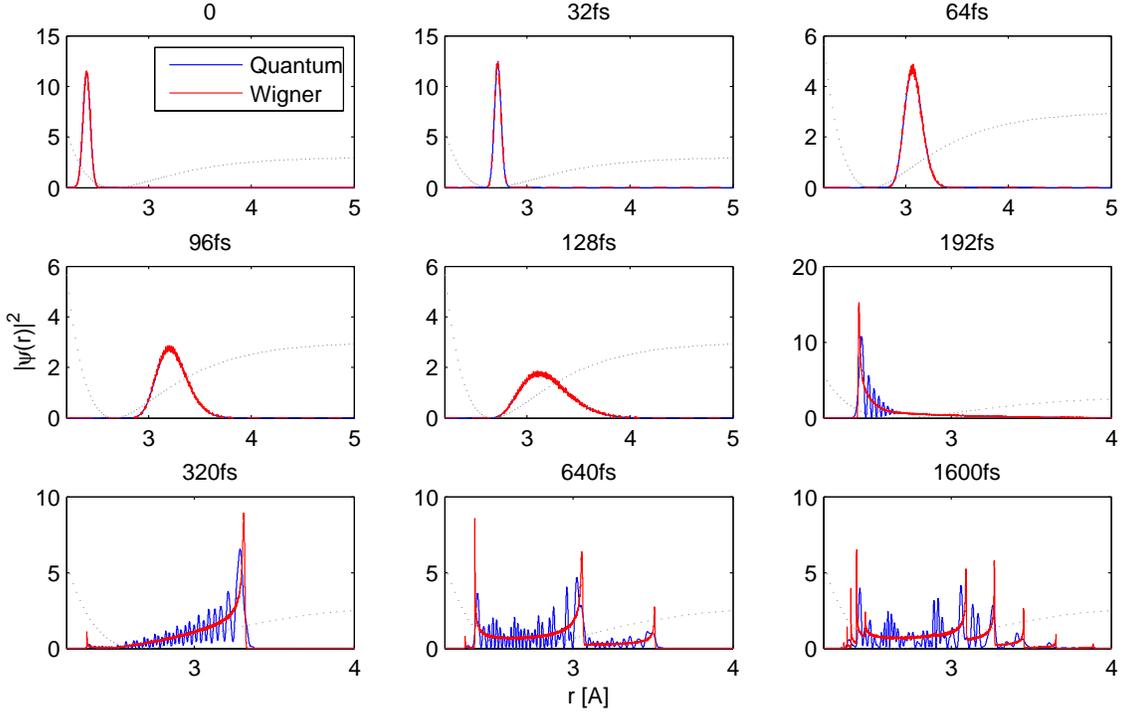}
\caption{Quantum propagation and quasi-classical evolution of the position density for a Morse potential which corresponds 
to the electronic ground state of $I_2$. The initial state is a Gaussian wavepacket centered at\cite{WTSGGM01}
 $r=2.4$\AA.}\label{fig:evolution}
\end{figure*}

Lastly, we  compare the quantum mechanical references $|(U_t\psi)(r)|^2 = |\psi_t(r)|^2$ with the approximative position densities $P_t(r)$.
As in Ref.\cite{TW04,WTSGGM01} we show snap shots for different times, see  
Figure~\ref{fig:evolution}. Up to time $128$fs, both position densities are almost indistinguishable. 
But also afterwards, even up to $1600$fs,  $P_t(r)$ represents a decent mean position density and 
displays the localization areas and strong  peaks of the quantum mechanical position density better than expected. 

\begin{figure}[h!]
\includegraphics[width=7.5cm]{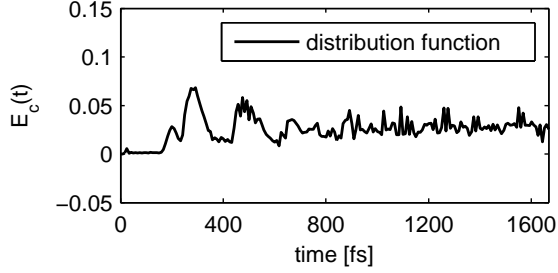}
\caption{Evolution of the maximal deviation (\ref{eq:Linfty_error_pos}) of the cumulative position distribution functions associated with $P_t(r)$ and $|\psi_t(r)|^2$.
\label{fig:pos_cumulate_error}}
\end{figure}

To substantiate these observations, we introduce two different error measures, namely the integrated difference
\begin{equation}\label{eq:L1_error_pos_dens}
E_1(t) = \int_{0}^\infty \Big|   |\psi_t(r)|^2 - P_t(r) \Big|dr
\end{equation}
and the maximal deviation of the cumulative distribution functions
\begin{equation}\label{eq:Linfty_error_pos}
E_c(t) = \sup_{x\ge0} \Big| \int_{0}^x \left( |\psi_t(r)|^2~- P_t(r) \right) dr \Big|.
\end{equation}
We always have
$$
E_c(t) \le E_1(t).
$$
In our numerical experiments, however, the cumulative error is considerably smaller than the integrated one:  
\begin{figure}[h!]
\includegraphics[width=7.5cm]{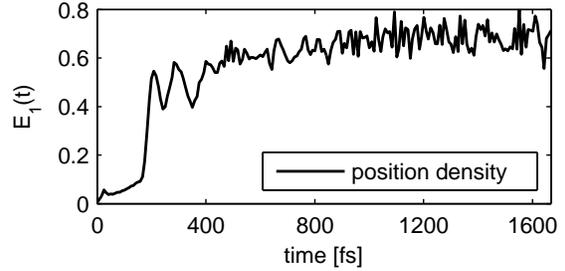}
\caption{Evolution of the integrated difference (\ref{eq:L1_error_pos_dens}) of the quasi-classical position density $P_t(r)$ and the reference density $|\psi_t(r)|^2$.}\label{fig:pos_dens_error}
\end{figure}
Figure~\ref{fig:pos_dens_error} shows that the integrated difference $E_1(t)$  stays small only until $170$fs and detoriates
afterwards, illustrating the limitations of quasi-classical approximations as previously discussed in the literature\cite{DH84,WTSGGM01,TW04}.   
By contrast,  Figure~\ref{fig:pos_cumulate_error} displays the much smaller deviation of the cumulative distribution functions~$E_c(t)$, which stays below $\sqrt\e \approx 0.11$ also for longer times, see \S\ref{sec:LSC} and Appendix~\ref{app:heur}.


\subsection{Henon--Heiles dynamics in higher dimensions}\label{sec:numerics_henon}

We present computations with  confined Henon--Heiles potentials\cite{MMC90,RM00} 
in dimensions $2$ to $32$ which illustrate the performance
of the quasi-classical algorithm in moderately high-dimensional situations.  
Henon--Heiles systems have been previously simulated by different methods as the multiconfiguration time-dependent Hartree method (MCTDH)\cite{MMC90,RM00,NM02}, semiclassical initial value representations\cite{B99,WMM01,TW04} and coupled coherent states\cite{SC04}. These studies have mostly aimed at the autocorrelation function $c_t = \langle\psi\mid U_t\mid\psi\rangle$ and its Fourier transform. For quasi-classical approximations only the modulus 
\[
(c_t^*c_t)^{1/2} = \(\lw U_t \psi \mid\psi\rw \lw \psi \mid  U_t \psi\rw\)^{1/2}
\] and not the complex number $c_t$ 
is computable.

\subsubsection{The model system}
\begin{figure}[ht!]
\includegraphics[width=7.5cm]{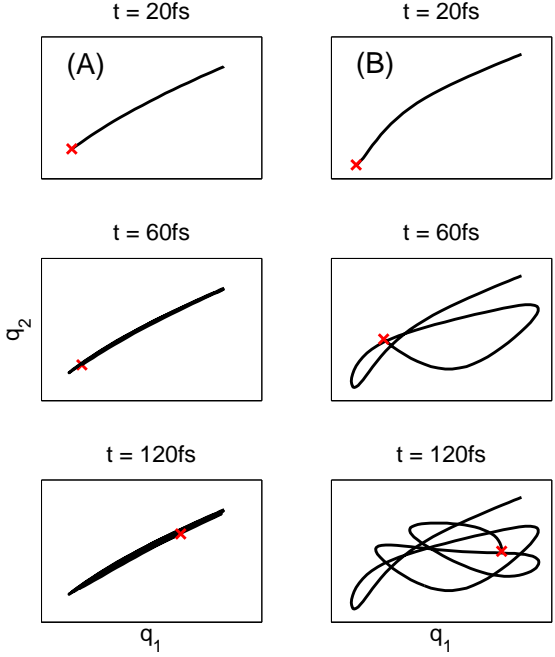}
\caption{Evolution snapshots of the expected position trajectories for initial 
data (A) and (B) on the left and right hand side, respectively. The red cross indicates the current position.\label{fig:henon_trajectory}}
\end{figure}

We investigate the dynamics of a 
hydrogen atom on a $d$-dimensional PES represented by the Henon--Heiles potential
\begin{equation*}
\frac{m\o^2}2 \sum_{j=1}^d q_j^2 + \sigma \sum_{j=1}^{d-1}(q_j^2q_{j+1} - \tfrac13 q_{j+1}^2)
\end{equation*}
with\cite{NM02} $m=1837m_e$, 
$\sigma = 0.0072 E_h a_0^{-3}$, and $m\o^2 = 0.0248 m_e ({\rm a.t.u.})^{-2}$.
Rescaling space according to $q\mapsto q/\overline q = \sqrt{0.0248}\,q$, we obtain the Schr\"odinger operator in semiclassical scaling 
\begin{eqnarray*}
\lefteqn{\widehat H =}\\
&& -\tfrac{\e^2}2 \Delta + \tfrac12 \sum_{j=1}^d q_j^2 +  1.8436 \sum_{j=1}^{d-1}(q_j^2q_{j+1} - \tfrac13 q_{j+1}^2)
\end{eqnarray*}
with $\e=0.0037$. 
Due to the large coupling constant $1.8436$, the MCTDH calculations\cite{NM02} for this Hamiltonian have employed 
complex absorbing potentials. Following Refs.\cite{MMC90,RM00}, we do not add a complex absorber but a quartic confinement that prevents phase space trajectories from escaping to infinity. Our modified Henon-Heiles potential  reads
\begin{align}\label{eq:henonhe}
V_d(q) = &\nonumber\tfrac{1}2 \sum_{j=1}^d q_j^2 + 1.8436\sum_{j=1}^{d-1}(q_j^2q_{j+1} - \tfrac13 q_{j+1}^2)\\
& + 0.4 \sum_{j=1}^{d-1}(q_j^2 + q_{j+1}^2)^2.
\end{align}
As considered previously\cite{NM02}, we investigate as initial data (A) the shifted harmonic ground state (\ref{eq:coherent_state}) of width $\Sigma=\Id$,
with initial position $q_{0,k}=0.0408$nm and initial momentum $p_{0,k}=0$ for all $k=1,\hdots,d$. In the rescaled
units, the shift equals $q_{0,k} \approx 0.1215\overline{q}$.
This choice leads to an almost periodic evolution of the expected positions, see Figure~\ref{fig:henon_trajectory}.
We contrast this setup by
computations for Gaussian initial  data (B) localized in $q_{0,k}= 0.3645\overline{q}$ for all $k=1,\ldots,d$. 
Showing a ball of wool for the position expectations, Figure~~\ref{fig:henon_trajectory} proves that the dynamics for these higher energy wave functions are less regular.

For both initial conditions, the total energy of the system grows with the dimension.

\subsubsection{The numerical setup}
In dimension $d=2$, we compare the approximative expectation values of the kinetic, potential, and total energies
obtained from the quasi-classical algorithm with reference data from a high resolution Strang splitting 
for the corresponding Schr\"odinger equation, see Table~\ref{tab_ref}. 
\begin{table}[ht!]
\centering
\begin{tabular}{c|c|c|c}
space area& Fourier modes &  time  & timesteps\\ \hline
 $[-5,5]\times[-5,5]$ & $2048 \times 2048$ & $[0,50]$  &  $5\cdot10^5$ \\
\end{tabular}
\caption{\label{tab_ref} Data of the reference solution of the vibrational Schr\"odinger equation for the 
$2$-dimensional Henon-Heiles system, generated by a split step Fourier solver on the time interval $[0,327{\rm fs}]$. }
\end{table}
Because of the time rescaling, the final time of $50/\e$ atomic time units equals $327$fs.
In dimensions $d>2$ we restrict 
ourselves to the comparison of the evolution of 
potential  energies, the preservation of the total energy and the computational effort. 

For all classical trajectories,
we used an symplectic integrator\cite{Y90} of order $8$ with time stepping $\tau=10^{-1}$.

\subsubsection{Energy expectation values}
\begin{figure}[ht!]
\includegraphics[width=7.5cm]{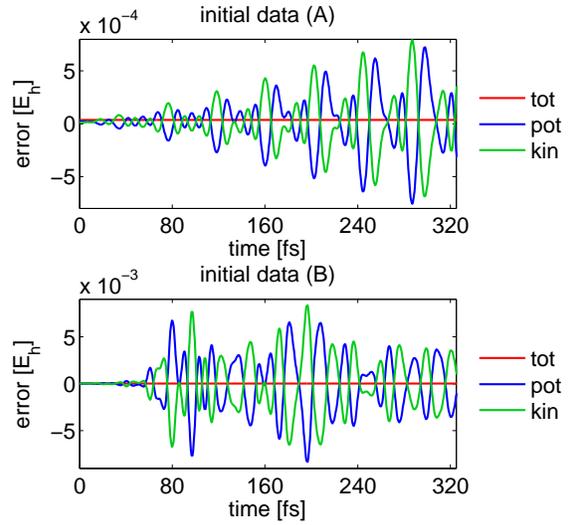}
\caption{Differences between the expectation values of the kinetic, potential, and total 
energies obtained from the quasi-classical algorithm with $N=2^{11}$ Sobol points and the references for the $2$-dimensional Henon-Heiles system.\label{fig:henon2d_long}}
\end{figure}

Figure~\ref{fig:henon2d_long} shows the error of the quasi-classically computed expectation values of the total, potential and kinetic energy in dimension $d=2$ with $2^{11}$ Sobol points. 
The errors are small but larger than the ones obtained for the iodine system in Figure~\ref{fig:morse_error} 
and differ for the two initial data on the long time scale of the simulation. 
In particular,  initial state (B) leads to less stable classical dynamics and a much faster local growth of errors than state (A).

\begin{figure}[h!]
\includegraphics[width=7.5cm]{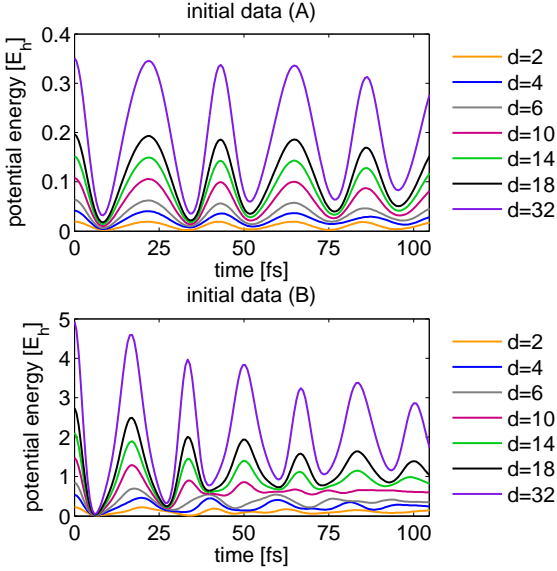}
\caption{Evolution of the expected potential energies computed by the quasi-classical algorithm in dimensions $4$ to $32$. 
}
\label{fig:henon_multid}
\end{figure}

In Figure~\ref{fig:henon_multid}, we present the time-evolution of the potential energy in dimensions $d=2,\ldots,32$ up to the shorter time $104\text{fs}$ where we used $2^{12}$ Sobol points for each of the calculations.
The results show highly regular oscillations for setup (A), and slightly damped dynamics in the case of  initial data (B).

\begin{figure}[h!]
\includegraphics[width=7.5cm]{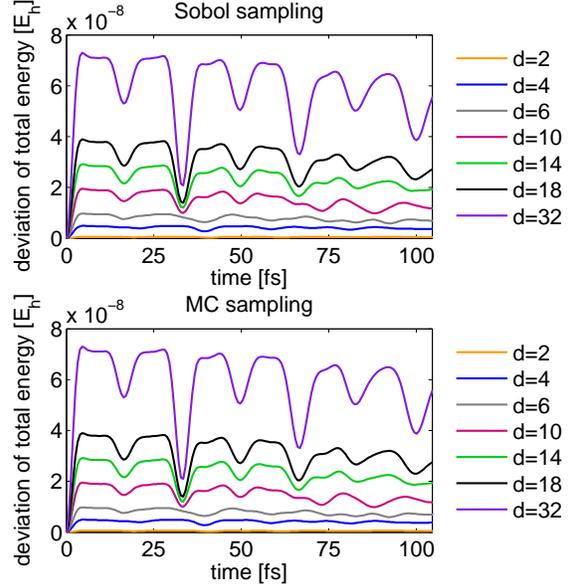}
\caption{Deviation from the initial value of the total energy by the quasi-classical algorithm for initial data (B)
in dimensions $2$ to $32$. }\label{fig:henon_deviation}
\end{figure}

Also the total energy deviation in Figure \ref{fig:henon_deviation} has regular oscillations, which are bounded by 
$8\cdot10^{-8}$ as expected for a symplectic eighth order time discretization with step size $\tau=10^{-1}$. We note, that the results for a Monte Carlo sampling with $2^{14}$ normally distributed points and those for a Quasi-Monte Carlo sampling with $2^{12}$ Sobol points of the initial Wigner function are almost indistinguishable.



The computational times grow moderately with the dimension, reaching less than 20 seconds for the 32-dimensional case, see Table~\ref{tab_henon}. 
\begin{table}[ht!]
\centering
\begin{tabular}{c||c|c|c|c|c|c|c}
$d$ & 2 & 4 & 6 & 10 & 14 & 18 & 32\\ \hline
comp. time & 0.7s & 1.3s & 2.0s & 5.9s & 7.7s & 9.7s & 19.8s
\end{tabular}
\caption{\label{tab_henon} Computational time (in seconds) for the propagation of $N=2^{12}$ Sobol points for the 
$d$-dimensional Henon-Heiles system up to  $104\text{fs}$. }
\end{table}

\subsubsection{Bath energy}
Lastly, we revisit the dynamics for  the $32$-dimensional potential with slightly different initial data. 
As in Ref.\cite{NM02}, we view the last but four coordinates $q_5,\hdots ,q_{32}$ as bath degrees of freedom, and use 
the initial harmonic ground state with displacement $q_{0,k} \approx 0.1215\overline{q}$ only for the system 
coordinates $k=1,\hdots,4$, while the bath degrees of freedom are localized at the origin.

\begin{figure}[h!]
\includegraphics[width=7.5cm]{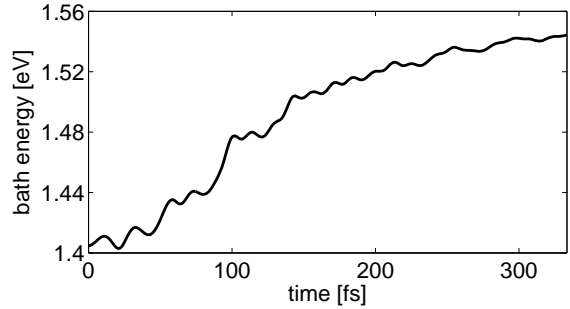}
\caption{Quasi-classical evolution of the expected bath energy defined in~(\ref{eq:bath_en}). }\label{fig:henon_bath}
\end{figure}

We are interested in the evolution of the expectation value of the bath energy
\begin{align}
\widehat{H_B} =\nonumber &\tfrac{1}2 \sum_{j=5}^{32}(-\e^2 \partial_{q_j}^2 + q_j^2) + 0.9218 q_4^2q_5\\
&\label{eq:bath_en}+ 1.8436\sum_{j=5}^{31}(q_j^2q_{j+1} - \tfrac13 q_{j+1}^2),
\end{align}
where the bath coupling term $1.8436q_4^2q_5$ has been divided equally between the system and the bath Hamiltonian.
The quartic confinement or rather the missing complex absorbing potential in $V_{32}$ does not allow for a quantitative comparison of the results in Figure~\ref{fig:henon_bath} with the MCTDH calculations\cite{NM02}. Nevertheless,
the qualitative structure and the range of the dynamics agree well. The bath energy computation for Figure~\ref{fig:henon_bath} used $2^{12}$ Sobol points and took $63$ seconds.

\section{Conclusion}\label{sec:con}

We have related quasi-classical approximation schemes as the linearized semiclassical initial value representation (LSC-IVR) and the Wigner phase space method to Egorov's theorem. Depending on the quantity of interest, the error estimates may depend on the initial data: For the computation of typical expectation values, only normalized initial wave functions with $\langle\psi|\psi\rangle =1$ are required for an error of the order $\e^2$. For the computation of position densities and transition probabilities, however, higher order derivatives of the inital Wigner function influence the accuracy of the approximation, such that for localized initial data the accuracy drops down to $\sqrt\e$.

Our numerical experiments for a Morse model of diatomic iodine and for  confined Henon--Heiles systems in various dimensions have illustrated the theoretical results but have also shown persistence on longer time scales than expected. The computational times are in the range of seconds.

The numerical computation of the approximation error's proportionality factor $E(V,B,\Phi_t)$ can be achieved by ordinary differential equations involving higher order derivatives of the potential $V$ and the observable $B$. So far, these factors have successfully been computed for two-dimensional torsional dynamics\cite{GL14}, and the application to more demanding test systems seems to be a natural continuation of the research presented here.

\section{Acknowledgments}

This research was supported by the German Research
Foundation (DFG), Collaborative Research Center SFB-TRR 109, and the graduate
program TopMath of the Elite Network of Bavaria.

\appendix
\section{Proof of Egorov's theorem}\label{proof}

Egorov's theorem has a simple proof\cite{BR02}, whose key element is the asymptotic expansion of the commutator of Weyl quantized observables  in even powers of $\e$,
\begin{eqnarray*}
\frac{i}{\e}\!\left[\widehat A,\widehat B\right] &=& \frac{i}{\e}\!\left(\widehat A \widehat B - \widehat B \widehat A\right)\\
&\sim & \sum_{k=0}^\infty \left(\frac{\e}{2i}\right)^{2k} \widehat{\{A,B\}}_{2k+1},
\end{eqnarray*}
where the $k$th order Poisson bracket is defined according to
$$
\{A,B\}_k = \sum_{|\alpha+\beta|=k} \frac{(-1)^{|\beta|}}{\alpha!\beta!} \left(\partial_q^\alpha \partial_p^\beta B\right) \left(\partial_q^\beta \partial_p^\alpha A\right).
$$
One argues as follows:
\begin{eqnarray*}
\lefteqn{U_{-t}\,\widehat B \,U_{t} - \widehat{B\circ\Phi_t}}\\ 
&=&
\int_0^t \frac{d}{ds} \left(U_{-s}\widehat{B\circ\Phi}_{t-s}U_{s}\right) ds\\
&=&
\int_0^t U_{-s} \left(\tfrac{i}{\e}[\widehat H,\widehat{B\circ\Phi}_{t-s}] -\partial_t \widehat{B\circ\Phi}_{t-s}\right) U_{s} ds\\
&\sim& \sum_{k=1}^\infty \left(\frac{\e}{2i}\right)^{2k} \int_0^t U_{-s}\, \op({\{H,B\circ\Phi_{t-s}\}}_{2k+1}) U_s \,ds,
\end{eqnarray*}
with $\op(A) = \widehat{A}$, since
$$
\partial_t(B\circ\Phi_{t-s}) = \{H,B\circ\Phi_{t-s}\}_1.
$$
The second order term
$$
-\frac{\e^2}{4} \int_0^t U_{-s}\, \op(\{H,B\circ\Phi_{t-s}\}_{3}) \,U_s \,ds
$$
is expected to dominate the approximation error $E(V,B,\Phi_t)$ in Egorov's theorem (\ref{egorov}).

\section{Approximating position densities}\label{pd}

In contrast to the LSC-IVR approximation error, which is uniform over all initial states with $\langle\psi\mid\psi\rangle=1$, the accuracy of the 
quasi-classical computation of position densities\cite{SM99} by a combination of Egorov's theorem with the Fourier inversion formula depends on the initial state: The error of (\ref{eq:pos_dens_app}) is
$$
\e^2 \langle\psi\mid \int E(V,B_\eta,\Phi_t) d\eta \mid \psi\rangle.
$$
The dominant part of this term contains third order derivatives of the potential $V$ and the observable $B_\eta$, that is, 
\begin{eqnarray*}
\lefteqn{\int_0^t \int \langle U_s\psi\mid \op(\partial_q^3 V \partial_p^3 (B_\eta\circ\Phi_{t-s})\mid U_s\psi\rangle d\eta ds}\\ 
&=&\int_0^t \int W_{s}(z) \,\partial_q^3 V(q)\, \partial_p^3 (e^{-i\eta\cdot(q_{t-s}-r) }) dz d\eta ds\\
&=&-\int_0^t \int (\partial_p^3 W_{s})(\Phi_{s-t}(z)) \,(\partial_q^3 V)(q_{s-t})\\
&& \qquad\qquad e^{-i\eta\cdot(q-r) } dz d\eta ds,
\end{eqnarray*}
where $W_s$ denotes the Wigner function of the time-evolved wave function $U_s\psi$. This implies, that the error is not uniform over all initial wave functions $\psi$ with $\langle\psi|\psi\rangle=1$, but crucially depends on third derivatives of its time-evolved Wigner function. 

\section{Heuristics for Wigner derivatives}\label{app:heur}

We present a heuristic argument, explaining the considerable difference between the integrated error measure $E_1(t)$ and the cumulative measure $E_c(t)$ proposed in \S\ref{sec:posdens}. 

If the initial state~$\psi$ is a 
vibrational Gaussian wavepacket~(\ref{eq:coherent_state}), then there are $\Sigma$-dependent complex numbers $c_{m,n}$ such that
$$
\partial_p^3 W_\psi = \e^{-3/2} \sum_{|(m,n)|\le 3} c_{m,n} \,W(\psi_m,\psi_n),
$$
where
\begin{eqnarray*}
\lefteqn{W(\psi_m,\psi_n)(q,p) =}\\
&&  (2\pi\e)^{-d} \int \psi_m(q-\tfrac12y)\psi_n^*(q+\tfrac12y) e^{i p\cdot y/\e} dy
\end{eqnarray*}
denotes the joint Wigner function of two generalized coherent states\cite{H98,LT14} $\psi_m$ and $\psi_n$. We approximate
\begin{eqnarray*}
\lefteqn{(\partial_p^3 W_s)(\Phi_{s-t})(z) \approx}\\
&& \e^{-3/2} \sum_{|(m,n)|\le 3} c_{m,n} \,W(U_t\psi_m,U_t\psi_n)
\end{eqnarray*}
and obtain
\begin{eqnarray*}
\lefteqn{|(U_t\psi)(r)|^2 - P_t(r) \approx}\\
&& \sqrt{\e} \,F(V,\Phi_t)\sum_{|(m,n)|\le 3}(U_t\psi_m)(r) \,(U_t\psi_n)(r)^*,
\end{eqnarray*}
where $F(V,\Phi_t)$ depends on the potential $V$ and the flow $\Phi_t$. This implies for the two error measures
\begin{eqnarray*}
\lefteqn{E_1(t) \approx\sqrt\e\, |F(V,\Phi_t)| }\\
&& \int_0^\infty \Big| \sum_{|(m,n)|\ge 3}(U_t\psi_m)(r)\, (U_t\psi_n)(r)^* \Big| dr 
\end{eqnarray*}
and
\begin{eqnarray*}
\lefteqn{E_c(t) \approx\sqrt\e\, |F(V,\Phi_t)| }\\
&& \sup_{x\ge0}\Big|  \sum_{|(m,n)|\le 3} \int_0^x (U_t\psi_m)(r)\, (U_t\psi_n)(r)^* dr\Big|.
\end{eqnarray*}
The decisive difference between the two error measures is therefore, that $E_1(t)$ depends on the integrated modulus of products of excited coherent states, whereas $E_c(t)$ sees the modulus of their cumulative overlap.


\providecommand{\noopsort}[1]{}\providecommand{\singleletter}[1]{#1}%

\end{document}